**Novel voltage signal at proximity effect induced superconducting transition temperature in gold nanowires**

Jian Wang[1-5,*], Junxiong Tang[1,†], Ziqiao Wang[1,4], Yi Sun[1], Qing-Feng Sun[1,2,5], Moses H. W. Chan[4,*]

[1]*International Center for Quantum Materials, School of Physics, Peking University, Beijing 100871, China*

[2]*Collaborative Innovation Center of Quantum Matter, Beijing, China*

[3]*State Key Laboratory of Low-Dimensional Quantum Physics, Department of Physics, Tsinghua University, Beijing 100084, China*

[4]*Department of Physics, The Pennsylvania State University, University Park, Pennsylvania 16802-6300, USA*

[5]*CAS Center for Excellence in Topological Quantum Computation, University of Chinese Academy of Sciences, Beijing 100190, China*

[*] *Corresponding authors: jianwangphysics@pku.edu.cn; mhc2@psu.edu*

[†] *Present address: Department of Physics, University of Colorado Boulder, CO 80309, USA*

We observed a novel voltage peak in the proximity effect induced superconducting gold (Au) nanowire while cooling the sample through the superconducting transition temperature. The voltage peak turned into dip in the warming process. The voltage peak (or dip) was found to be closely related to the emergence (vanishing) of the proximity induced superconductivity in the Au nanowire. The amplitude of the voltage signal depends on the temperature scanning rate and it cannot be detected when the temperature is changed slower than 0.03 K/min. This transient feature suggests the non-equilibrium property of the effect. The voltage peak could be understood by Ginzburg-Landau model as a combined result of the emergence of Cooper pairs with relatively lower free energy in W contact and the non-equilibrium diffusion of Cooper pairs and quasiparticles.

**proximity effect induced superconductivity, Au nanowire, non-equilibrium process**

# 1 Introduction

With wide applications from superconducting quantum interference devices (SQUIDs) [1, 2] to quantum computation [3, 4], superconductivity in nanoscale systems has attracted great deal of attention in condensed matter physics and nanotechnology [5-10]. Low dimensional superconducting systems exhibit behavior different from bulk superconductors. For example, fluctuation effects [11-13] become important and cannot be neglected. Moreover, superconducting proximity effect [14] plays a unique role in nanoscale systems. To be specific, normal material near the interface becomes superconducting due to the non-local property of Cooper pairs, and simultaneously superconductivity at the superconductor side is depressed near the interface [15]. Recent studies reveal proximity effect induced phenomena in ferromagnet/ superconductor [16-19], graphene/ superconductor [20] and topological insulator/ superconductor [21, 22] hybrid or hetero-nanostructural systems. Interestingly, possible experimental evidences of Majorana zero modes have been reported in proximity effect induced superconducting InAs nanowire and ferromagnetic chain with strong spin-orbit coupling [23, 24]. Thus, careful studies of the physical properties on other strong spin-orbit coupling nanowires with superconducting contacts are highly desirable.

In this paper, we report the detection of current-independent voltage signal in spin-orbit coupling Au nanowire/ superconducting W nanostrip hybrid structure. A novel voltage peak of ~ 10 μV was detected while cooling the system across the proximity effect induced superconducting transition temperature ($T_c$) of the Au nanowire, whereas the voltage peak turned into dip in warming process. Moreover, there is a close connection between the amplitude of voltage peaks (dips) and the temperature scanning rates, implying non-equilibrium properties of the voltage signal. We proposed a simplified model considering the emergence of Cooper pairs in W contact and the non-equilibrium diffusion of Cooper pairs to describe the exotic voltage peak.

# 2 Experimental observations

Our single crystalline gold nanowires with diameter of 70 nm were prepared by electrodeposition [25, 26]. Figure 1(a) shows a typical scanning electron microscope (SEM) image of individual Au nanowire on Si substrate with insulating $Si_3N_4$ layer contacted by two superconducting W compound electrodes in the middle and two normal Pt electrodes at the ends. The superconducting transition temperature of the focused ion beam (FIB) deposited W compound strip is around 5 K [27, 28]. Standard four probe measurements with two Pt current electrodes and two W voltage electrodes separated by 1.49 μm reveal a proximity effect induced superconducting transition from $R_n \sim 65$ Ω to less than 30 Ω at $T_c \sim 4$ K [Figure 2(c)]. The resistance does not reduce to zero, since the effective length between two voltage electrodes is longer than the proximity induced superconducting coherent length in the Au nanowire. Therefore, only the parts of the Au nanowire near the W electrodes are superconducting at low temperature while the center part of the nanowire remains normal.

The configuration for the "non-local" voltage measurements are shown in Fig 1(b). The current was applied on the nanowire via the two electrodes on the left side, while the voltage signal was detected across the other two electrodes (one Pt and one W) on the right. A voltage peak with magnitude of ~ 8 μV could be clearly identified around 4 K (black line in Figure 1(c)) when the system was cooling down from 6 K to 2 K at a rate of 0.1 K/min, whereas the voltage peak turns into dip (red line in Figure 1(c)) with almost the same magnitude and width while warming up from 2 K to 6 K. The voltage signal could not be an accidental error, since each data point in the figure is the average value of 25 measurements.

Control experiments were carried out to determine the influence of the excitation current applied on the left side of the Au nanowire. As shown in Figure 1(c) and 1(d), although the applied current has significantly different magnitude (5 nA and 100 nA, respectively), the difference of the voltage signal is rather small. Moreover, even if the direction of the excitation current is inverted [Figure 1(e) and 1(f)], the properties of the detected voltage peaks and dips remain almost unchanged. We even found that the voltage peak can still be detected in the absence of applied current on the nanowire [Figure S1]. These results indicate that the voltage peaks and dips detected on the right side of the Au nanowire are independent on the excitation current applied on the left. This property distinguishes the non-local voltage signals detected in our experiments from early studies [29-33]. In those measurements, the excitation current is important to the detected non-local voltage signal. In ref. [29-31], a drive current is required to induce charge imbalance between the electron-like branch and the hole-like branch, which is necessary for the pair-quasiparticle potential difference. Drive current is also necessary when the non-local voltage is induced by crossed Andreev reflection or elastic co-tunnelling [32, 33]. However, such a drive current is irrelevant to the "non-local" voltage detected here. Besides, two voltage electrodes in our devices are attached to "normal" Au nanowire, while in those early studies [29-33] the voltage electrodes are all attached to superconductors. Therefore, a different mechanism is needed to understand the "nonlocal" voltage signal shown in Figure 1.

Since the voltage peak (or dip) appears at temperature (~ 4.0 K) slightly lower than the zero resistance $T_c$ of the W electrode (~ 4.7 K) [28], the peak (or dip) is likely to have its origin from the emergence (or vanishing) of the proximity effect induced superconductivity in Au nanowire near the W electrode. Figure 2(a) displays the "non-local" voltage signals detected while cooling the sample at magnetic field of 0, 1 T, 2 T, 4 T and 6 T, respectively. The voltage peak broadens and splits into several smaller peaks and moves to lower temperature with increasing magnetic field. The splitting behavior is possibly due to the fact that vortices divide the nanowire into several separated superconducting regions. The position and width of the peak ($T^* \pm \Delta T$) at corresponding magnetic field $B^*$ was plotted at the bottom with an error bar in Figure 2(a) and extracted for empirical fitting $B^*(T) = B^{*0}[1-(T/T^{*0})^2]$ as shown in Figure 2(b). For comparison, standard four probe measurements were carried out to investigate the evolution of proximity effect induced superconductivity with increasing magnetic field in Au nanowire. The excitation current of 500 nA was applied at the two ends of the nanowire through the Pt electrodes, while the voltage signal was detected via the two W leads in the middle. Transition temperature range ($T_c \pm \Delta T$) for each magnetic field $B_c$ was labeled with an error bar in Figure 2(c) and fitted by the empirical formula $B_c(T) = B_c^0[1-(T/T_c^0)^2]$ in Figure 2(d). The fitting parameter for $B^*(T)$ ($T^{*0} = 3.90 \pm 0.05$ K, $B^{*0} = 6.3 \pm 0.4$ T, see Figure 2(b)) is consistent with that for $B_c(T)$ ($T_c^0 = 4.0 \pm 0.1$ K, $B_c^0 = 5.8 \pm 0.4$ T, see Figure 2(d)), demonstrating the close relation between the voltage signal and the proximity effect induced superconducting transition.

We then investigated the evolution of the voltage signals at different temperature scanning rate [Figure 3]. Voltage peak (or dip) could not be detected at a scanning rate of 0.01 K/min [Figure 3(a)]. A smaller peak of 3.5 μV appears when the scanning rate is increased to 0.03 K/min but the dip cannot be resolved from the noise [Figure 3(b)]. As the scanning rate is increased, the peak (or dip) amplitude increases and reaches ~ 10 μV at a rate of 0.2 K/min [Figure 3 (c)-(e)]. Noteworthily, time durations of the voltage signals are all on the order of one minute, although the temperature scanning rates vary significantly. This result implies a non-equilibrium relaxation process with long characteristic time, which makes it possible for the detection of the voltage signal via transport measurements. The long characteristic time was further confirmed by detecting the time evolution of the voltage signal while warming the sample up from 2 K to 3.95 K at 0.1 K/min and then keeping it at 3.95 K [Figure 3(f)]. The voltage dip appeared and then declined in time scale of one minute (defined as the full width at half maximum of the voltage dip).

## 3 Physical model

We will consider below the possible origin of the observed voltage signal. Thermo-electromotive force or potential is one possible reason. Since the cooling (and warming) of the Au nanowire through the thin insulating $Si_3N_4$ substrate at low temperature is not effective because of low thermal conductivity of $Si_3N_4$ and large Kapitsa boundary resistance [34], the Au nanowire is cooled mainly via the electrodes contacting the wire by means of normal electrons. When the W lead is cooled into the superconducting state, its thermal conductivity is significantly reduced compared to the normal state and much smaller than the normal Pt lead. This effect can give rise to a transient temperature difference across the Au nanowire section between the W and the Pt voltage electrodes and induce a voltage peak or dip due to the thermo-electromotive force. However if this is the origin of the phenomenon, the voltage signal should appear at the superconducting transition temperature of the W lead, which is slightly above the proximity effect induced $T_c$ of the Au nanowire. More importantly, the voltage peak can still be detected in a Au nanowire where both voltage electrodes are superconducting W [Figure S2]. This allows us to rule out the thermos-electromotive force explanation.

We propose that the condensation of Cooper pairs in the W electrode and the non-equilibrium diffusion of Cooper pairs from W-Au junction into the Au nanowire could contribute to the voltage accumulation. Figure 4(a) illustrates the diffusion of Cooper pairs from the W-Au junction into the Au nanowire resulting in the emergency of the nonzero superconducting order parameter $\psi$ in the Au nanowire near the interface. To be simplified, we focus on non-equilibrium Cooper pair diffusion from the W electrode to the Au nanowire across the junction, and ignore the density gradient of Cooper pairs at each side of the W-Au interface. Andreev reflection [35] is also neglected since a majority of quasiparticles could enter the superconducting region without being reflected at $T \sim T_c$ [36]. The Cooper pair densities on the W and Au sides of the interface are labeled as $s_1$, $s_2$, respectively. $s_1$ and $s_2$ begin to be nonzero when the system is cooled down below $T_c$. In the absence of magnetic field and gradient, the Ginzburg-Landau theory gives free energy density $f_{SC} - f_N = \alpha|\psi|^2 + \frac{\beta}{2}|\psi|^4$ with $\alpha = A(T-T_c) = -Aut$, where $u$ is the temperature scanning rate and $t = 0$ is defined as the time when temperature is cooled to $T_c$. Cooper pairs in the W electrode generate at a rate $\frac{ds_1}{dt} = \frac{d|\psi|^2}{dt} = \frac{Au}{\beta}$. The proximity effect induced superconductivity in Au nanowire is achieved by Cooper pair diffusion from W across the interface, and the diffusion rate is assumed to be proportional to the difference of Cooper pair density between the two sides $s_1 - s_2$, namely $\frac{ds_2}{dt} = K(s_1 - s_2)$, and $\frac{ds_1}{dt} = \frac{Au}{\beta} - K(s_1 - s_2)$. Here, $K$ is the diffusion coefficient, and $K^{-1}$ is characteristic time scale of the diffusive process. Suppose $s_1, s_2 = 0$ at $t = 0$, we have the time evolution of $s_1$ and $s_2$

$$s_1 = \frac{Au}{2\beta}t + \frac{Au}{4\beta K}(1-e^{-2Kt}), \quad s_2 = \frac{Au}{2\beta}t - \frac{Au}{4\beta K}(1-e^{-2Kt}). \quad (1)$$

We assume that the normal state free energy density $f_N$ and the G-L parameter $\alpha$, $\beta$ of the Au nanowire are equal to that of the W electrode near the Au-W interface [37]. Then the difference of free energy between W and Au can be calculated as

$$f_W - f_{Au} = -Aut(s_1 - s_2) + \frac{\beta}{2}(s_1^2 - s_2^2) = \frac{-A^2u^2t^2}{4\beta Kt}(1-e^{-2Kt}) = \frac{-\alpha^2}{4\beta Kt}(1-e^{-2Kt}). \quad (2)$$

Here, $\frac{\alpha^2}{2\beta} = \frac{1}{2}N(0)\Delta^2(T)$ is the superconducting condensation energy, where $N(0)$ is the density of states at Fermi energy, and the temperature dependence of the superconducting gap $\Delta(T)$ near $T_c$ for BCS-type superconductors is approximately $\Delta(T) \approx 1.74(1-\frac{T}{T_c})^{1/2}\Delta(0)$. Finally we have the voltage accumulation between the superconductor and the normal metal

$$U = \frac{f_W - f_{Au}}{-Ne} = 0.76\frac{N(0)\Delta^2(0)}{Ne}\frac{u}{KT_c}(1-e^{-2Kt}) = 0.76\frac{N(0)\Delta^2(0)}{Ne}\frac{u}{KT_c}(1-e^{-2(KT_c/u)(1-T/T_c)}). \quad (3)$$

The temperature dependence of voltage in eq. (3) is depicted in Figure 4(b) to illustrate the peak rising while cooling across $T_c$. In fact, the emergence of Cooper pairs with relatively lower free energy in the W electrode contributes to the positive voltage signal in the cooling process, while in the warming process the signal turns to negative due to the breaking of Cooper pairs in the W electrode. Meanwhile, the diffusion of Cooper pairs and quasiparticles from the W electrode to the Au nanowire, which is induced by the Cooper pair density gradient and the electric potential difference, suppresses the rising of the voltage and results in a voltage drop. To quantitatively estimate the amplitude of the calculated voltage signal, we assume a carrier density $N \approx N(0) \times 2k_\mathrm{B}T$, $\Delta(0) \approx 1.76 k_\mathrm{B} T_c = 607 \mu eV$ ($T_c = 4.0$ K as the fitting parameter in Fig 2(d)), and characteristic time scale $K^{-1} \approx 1$ min, then $U \approx 10 \mu\mathrm{V} \times (1 - e^{-2(KT_c/u)(1-T/T_c)})$, if temperature scanning rate $u$ = 0.1 K/min. This result is consistent with the magnitude of the experimentally detected voltage peak at $u$ = 0.1 K/min. Moreover, the calculated $U$ is proportional to $u$, in accordance with our observations in Figure 3. When the scanning rate is as low as 0.01 K/min, the calculated peak amplitude is reduced to ~ 1 μV, which is comparable to the experimental noise amplitude and becomes unobservable. In addition, $U$ is proportional to the characteristic time scale $K^{-1}$. The detectable voltage signals benefit from the low diffusion coefficient $K$, which could be a result of the nanoscale contact area between the Au nanowire and the W nanostrip.

More interestingly, similar phenomena can also be observed in FIB deposited Pt compound nanostrip [38] with superconducting W contacts, but the magnitude of the voltage signal is much smaller [Figure S3]. The finding in the Pt nanostrip suggests the phenomenon may very well be 'universal' in many if not all superconductor-normal material nanostructures and deserves further experimental and theoretical investigations.

## 4 Conclusions

In summary, we detected exotic voltage peak (or dip) in Au nanowire - superconducting W electrode hybrid structure while cooling (warming) the sample across the proximity effect induced transition region. The voltage peak exhibits non-equilibrium characteristic with relatively long relaxation time of one minute. We modeled this phenomenon by Ginzburg-Landau theory as a combined result of the emergence of Cooper pairs in superconducting W electrode and the non-equilibrium diffusion of Cooper pairs and quasiparticles from the interface into the Au nanowire. The detection of voltage peak offers direct evidence that the free energy of a system will decrease after entering the superconducting state. Since the applied current is not necessary and the sign of the voltage signal depends on the increasing/decreasing of the temperature, our finding might offer a new way to detect superconductivity in nanostructures without a driving current.

*We thank Mingliang Tian and Meenakshi Singh for the helpful discussions. This work was financially supported by the National Basic Research Program of China (Grant No. 2017YFA0303300 and Grant No. 2013CB934600), the National Natural Science Foundation of China (Grant No. 11774008), the Open Research Fund Program of the State Key Laboratory of Low-Dimensional Quantum Physics (Grant No. KF201703), Tsinghua University, the Key Research Program of the Chinese Academy of Sciences (Grant No. XDPB08-1), and the Peking University President's Fund for Undergraduate Research (2013). The work at Penn State was supported by NSF grants (MRSEC) DMR-0820404 and DMR-1420620.*

## Supporting Information

The supporting information is available online at http://phys.scichina.com and . The supporting materials are published as submitted, without typesetting or editing. The responsibility for scientific accuracy and content remains entirely with the authors.

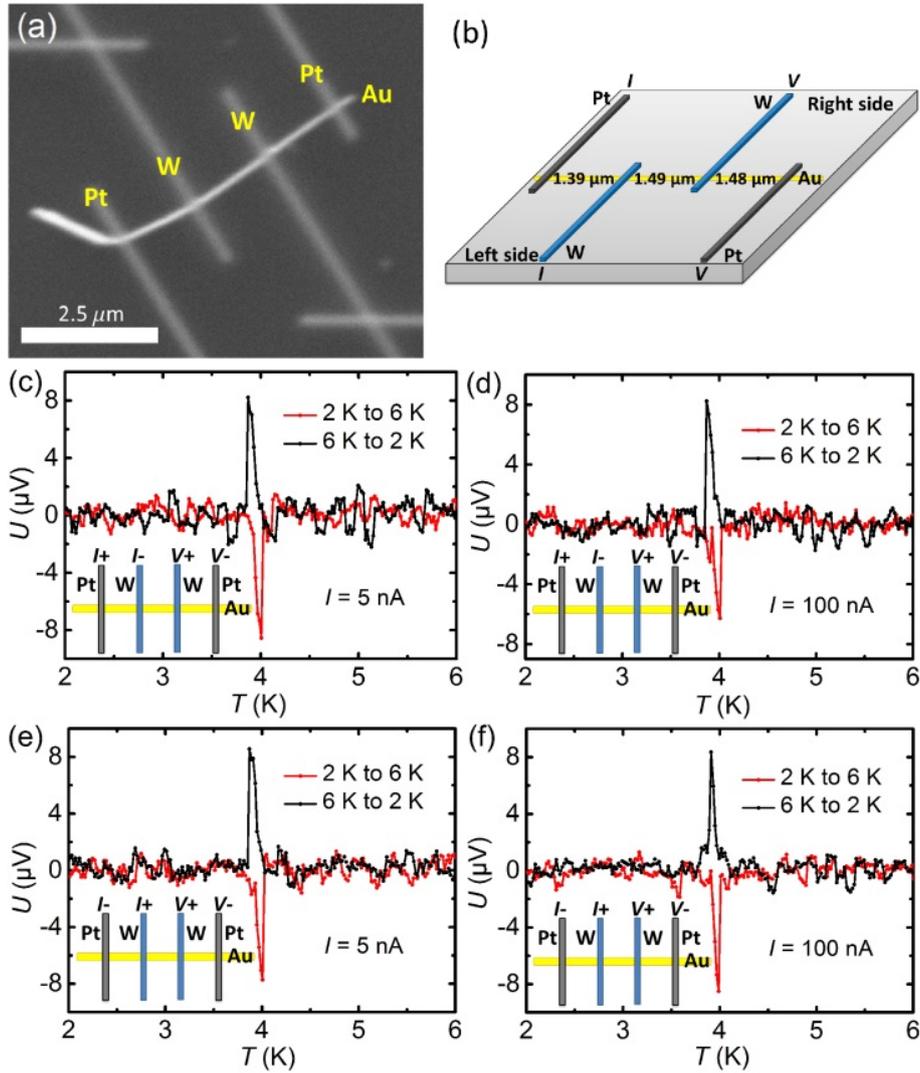

**Figure 1** (Color online) (a) A SEM image of Au nanowire contacted by two superconducting W compound electrodes in the middle and two normal Pt electrodes at the ends. (b) Schematic for non-local voltage measurement of Au nanowire. The excitation current was applied via the two electrodes at the left side, while the voltage signal was detected via the other two electrodes at the right side. (c) and (d) The detected non-local voltage signals with applied currents of (c) 5 nA and (d) 100 nA while cooling (black lines) or warming (red lines) between 2 K and 6 K at rate of 0.1 K/min. (e) and (f) Voltage signals were detected in similar conditions with (c) and (d), but with opposite current direction. Each data point in (c)-(f) is averaged from 25 measurements. Insets of (c)-(f): Configuration of electrodes.

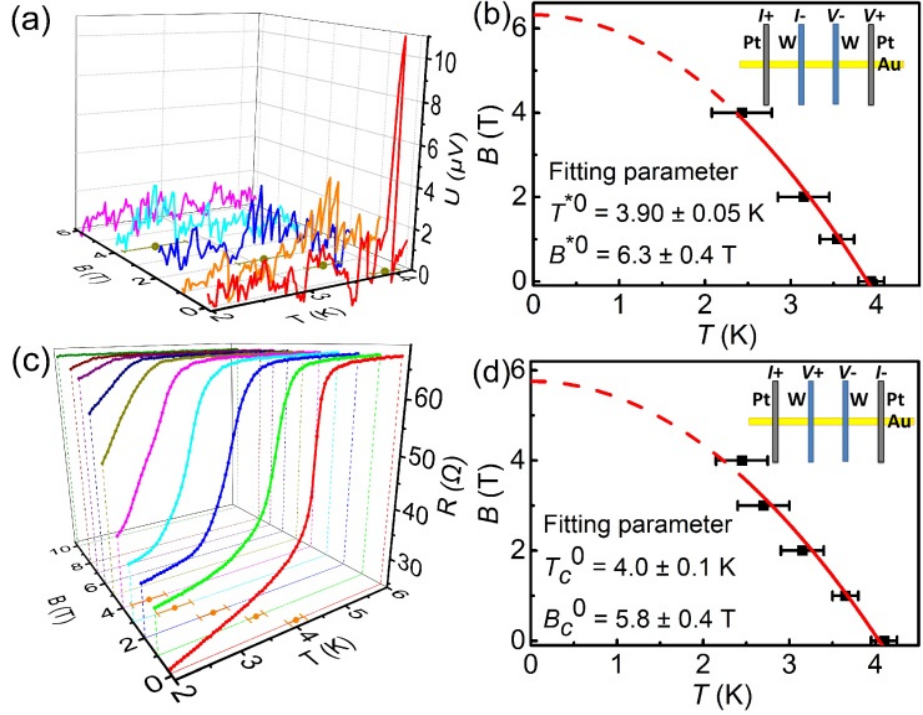

**Figure 2** (Color online) (a) The detected "non-local" voltage signals while warming the sample from 2 K to 6 K at rate of 0.1 K/min at magnetic fields of 0, 1 T, 2 T, 4 T, and 6 T, respectively. The position and width of voltage peak ($T^* \pm \Delta T$) at corresponding magnetic field $B^*$ were plotted at the bottom with error bar. (b) Fitting voltage peak data ($T^*$, $B^*$) by the empirical relation $B^*(T) = B^{*0}[1-(T/T^{*0})^2]$. (c) Standard four probe resistance as a function of temperature at various magnetic fields of 0, 1 T, 2 T, 3 T, 4 T, 5 T, 6 T, 7 T, 8 T, and 10 T, respectively. The transition range ($T_c \pm \Delta T$) was plotted at the bottom with error bars. $T_c$ is defined as the intersection of two black dashed lines, one is the linear extension of the superconducting transition drop and the other is extrapolated from the resistance tail. (d) Fitting superconducting transition data ($T_c$, $B_c$) by the empirical relation $B_c(T) = B_c^0[1-(T/T_c^0)^2]$. Insets of (b) and (d): Configuration of electrodes.

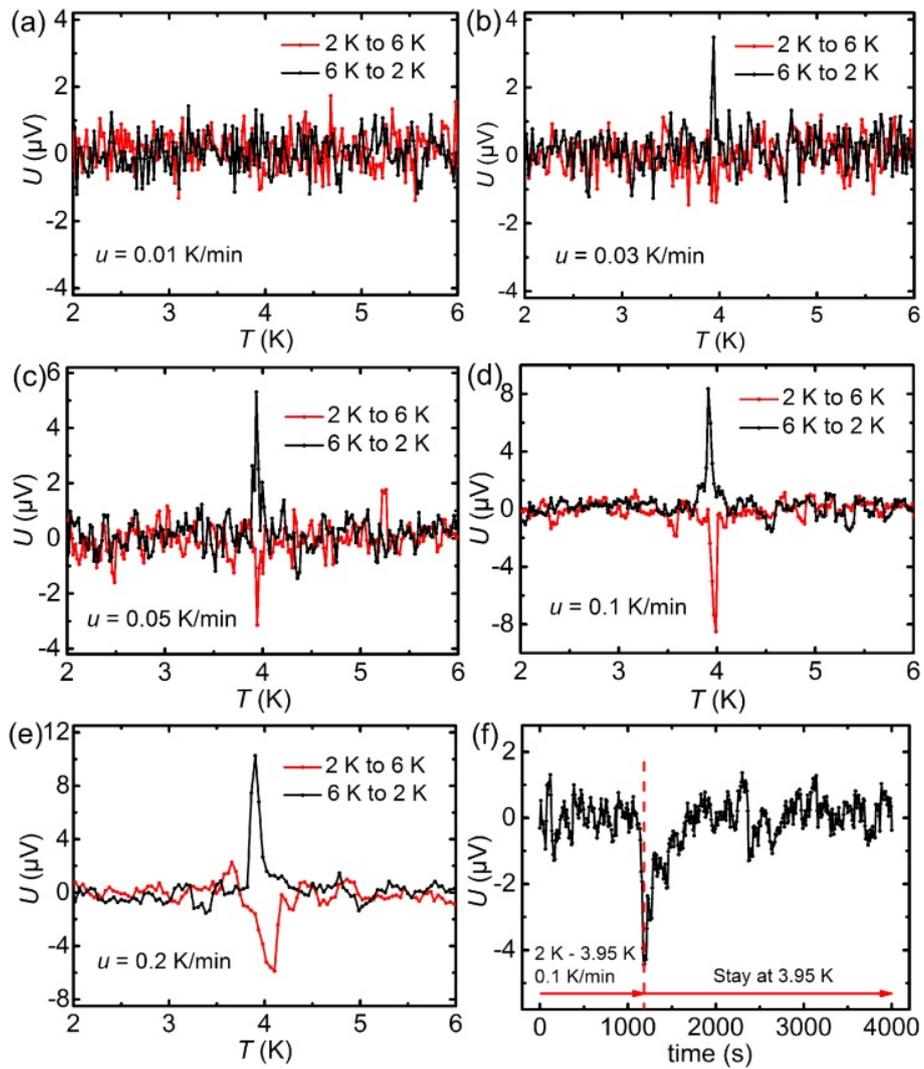

**Figure 3** (Color online) (a)-(e) The "non-local" voltage signals detected while cooling (black lines) or warming (red lines) between 2 K and 6 K at various temperature scanning rates *u* of 0.01, 0.03, 0.05, 0.1 and 0.2 K/min, respectively. The excitation current through the two leads on the left is 100 nA. (f) Time evolution of the voltage signal while warming the system up from 2 K to 3.95 K at 0.1 K/min and then keeping at 3.95 K.

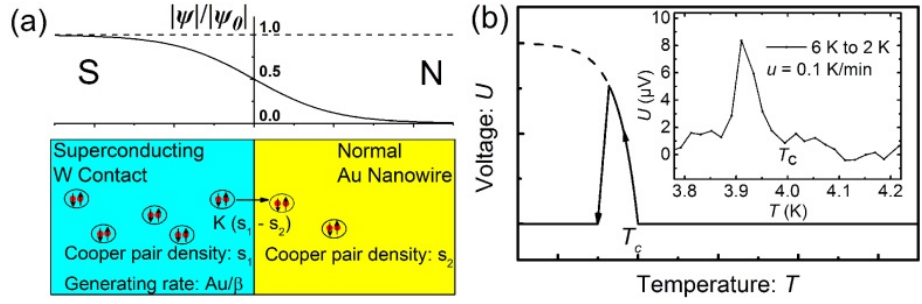

**Figure 4** (Color online) (a) Schematic of the non-local property of superconducting order parameter. The proximity effect induced superconductivity in Au nanowire is achieved by Cooper pair diffusion from the W lead across the interface. (b) Calculated voltage accumulation $U = -\dfrac{f_W - f_{Au}}{Ne}$ while cooling across $T_c$. The rising part of the voltage peak is depicted by eq. (3), while the voltage drop is a result of the diffusion of Cooper pairs and quasiparticles due to the electric potential difference. Inset: the magnified view of the voltage peak as a comparison with equation (3).

Supplemental Material for:

**Exotic voltage signal at proximity effect induced superconducting transition temperature in gold nanowires**

Jian Wang[1-4*], Junxiong Tang[1†], Ziqiao Wang[1,4], Yi Sun[1], Qing-Feng Sun[1,2], Moses H. W. Chan[4*]

[1]*International Center for Quantum Materials, School of Physics, Peking University, Beijing 100871, China*

[2]*Collaborative Innovation Center of Quantum Matter, Beijing, China*

[3]*State Key Laboratory of Low-Dimensional Quantum Physics, Department of Physics, Tsinghua University, Beijing 100084, China*

[4]*Department of Physics, the Pennsylvania State University, University Park, Pennsylvania 16802-6300, USA*

[*] Corresponding authors: jianwangphysics@pku.edu.cn; mhc2@psu.edu

[†] Present address: *Department of Physics, University of Colorado Boulder, CO 80309, USA*

1. Experimental data

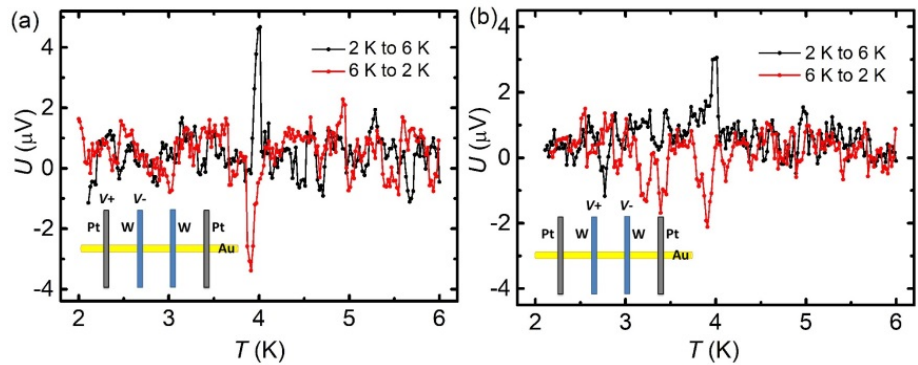

**Figure S1** The voltage signal can still be detected in the absence of applied current on the Au nanowire. (a) The voltage was detected across one normal Pt electrode and one superconducting W electrode. (b) The voltage was detected across two superconducting W electrodes. Amplitude of the voltage signal in (b) is smaller than that in (a).

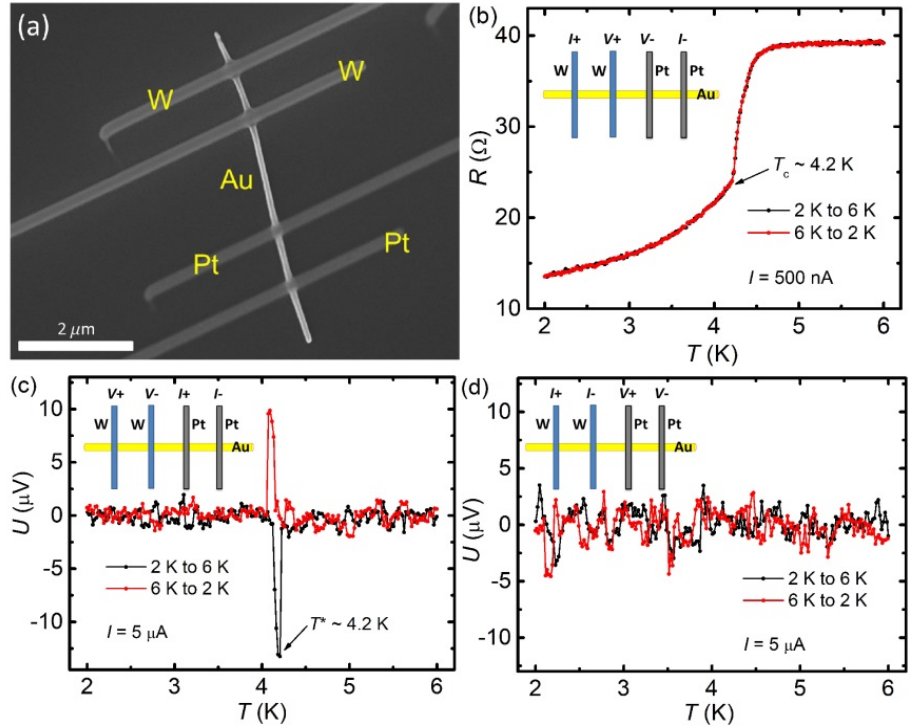

**Figure S2** (a) SEM image of another Au nanowire contacted by two superconducting W compound electrodes and two normal Pt compound electrodes. The scale bar is 2 μm. (b) Temperature dependent resistance curve of Au nanowire in (a), indicating proximity effect induced $T_c$ at ~ 4.2 K. (c) "Non-local" voltage signals are detected via two superconducting W leads at the proximity effect induced $T_c$. (d) "Non-local" voltage signals cannot be detected via two normal Pt leads. Temperature scanning rate in (c) and (d) is 0.1 K/min. Insets of (b)-(d): Configuration of electrodes.

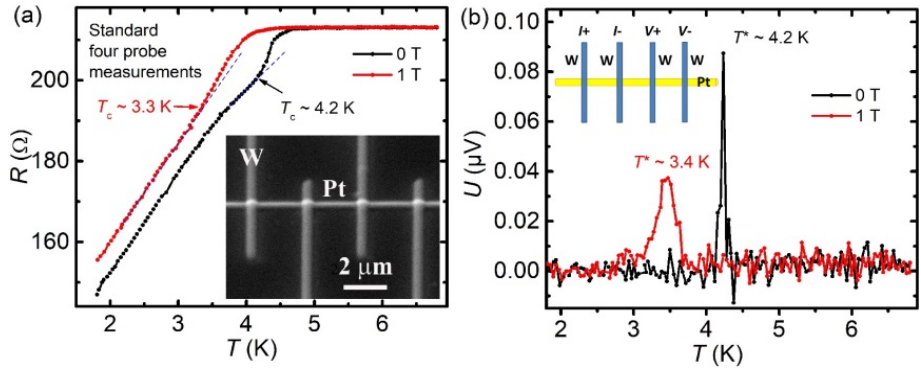

**Figure S3** (a) Temperature dependent resistance curves at magnetic field of 0 T and 1 T of Pt nanostrip contacted by four W electrodes. Proximity effect induced $T_c$ is marked by the $R(T)$ curve. Inset: SEM image of FIB deposited Pt compound nanostrip contacted by four superconducting W compound leads. (b) "Non-local" voltage signals are detected at the proximity effect induced $T_c(B)$. Temperature scanning rate is 0.1 K/min. Inset: Configuration of electrodes.

It is noteworthy to mention that we observed clear voltage signal at proximity effect induced $T_c$ in both Au and Pt nanowires, even if both the voltage electrodes are superconducting W leads. However, if the two voltage W leads on the nanowires are exactly the same, voltage signals at two contacts should counteract according to our model. Actually, in real situation, there is some difference between the two superconducting contacts. The difference in the diffusion coefficient $K$ at two contacts would contribute to different amplitude of the voltage signal, thus a net voltage signal could be detected. A more satisfactory theoretical understanding is highly desired.

2. Modified model including both Andreev reflection and direct diffusion of Cooper pairs.

There are two way for the Cooper pairs in the superconductor to enter a normal material. One is direct diffusion of Cooper pairs from the superconductor to normal material, the other is through Andreev reflection. Andreev reflection describes a physical process that an incident electron from the normal material is reflected back as a hole and a Cooper pair is injected into the superconductor. This process happens with a probability $K_A(T)$, and $K_A \approx 0$ when the temperature is close to Tc. That's why we can neglect Andreev reflection in the model described in the main text of the paper.

In the following, we calculate the voltage by modifying the model to include Andreev reflection at the electrode/Au nanowire interface together with direct diffusion of Cooper pairs. Note that for Andreev reflection, a Cooper pair from the superconductor splits into two normal electrons to the normal material, but for the direct diffusion of Cooper pairs, it stays as a Cooper pair inside the normal material. Let s1 and n1 be the density of superconducting and normal electrons in the superconductor, and let s2 and n2 be the density of superconducting and normal electrons in the normal metal respectively, we can show (see the left panel in Figure S4)

$$\frac{dn_1}{dt} = -\frac{Au}{\beta}$$

eq. (S1)

$$\frac{ds_1}{dt} = \frac{Au}{\beta} + bK(n_2 - n_1) - K(s_1 - s_2)$$

eq. (S2)

$$\frac{dn_2}{dt} = -bK(n_2 - n_1)$$

eq. (S3)

$$\frac{ds_2}{dt} = K(s_1 - s_2)$$

eq. (S4)

where in this case $K^{-1}$ is the characteristic time scale of diffusive process while cooling below Tc, b is the normalized conductance of Andreev reflection depending on temperature, voltage bias and interface barrier strength. For Andreev reflection through a perfect contact, b = 2. Finally we get

$$U_{sn} = \frac{f_s - f_n}{-Ne} = \frac{A^2 u^2 \left(1 - e^{-2Kt}\right)\left[1 - e^{-2Kt}(Kt + 1)\right]}{4\beta K^2 Ne} = \frac{A^2 u^2 \left(1 - e^{-2(T_c - T)}\right)\left[1 - e^{-2(T_c - T)}(T_c - T + 1)\right]}{4\beta K^2 Ne}$$

eq. (S5)

Voltage curve depicted eq. (S5) (see the right panel of Fig S4) is very similar to the curve in Fig 4(b) depicted by eq. (3) in the paper.

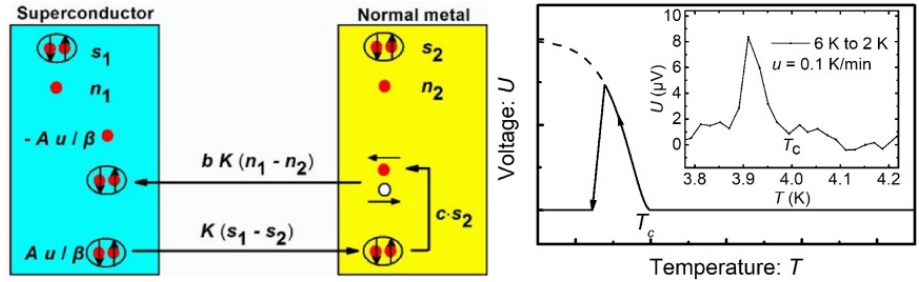

**Figure S4** (left) Diffusion of Cooper pairs considering both the Andreev reflection and direct penetration. (right) Voltage accumulation depicted by eq. (S5).